\documentclass[journal,12pt,onecolumn,twoside]{IEEEtranTCOM}
\normalsize

\usepackage{cite}

\ifCLASSINFOpdf
   \usepackage[pdftex]{graphicx}
\else
  \usepackage[dvips]{graphicx}
\fi

\usepackage[cmex10]{amsmath}
\usepackage{amssymb}

\usepackage{stfloats}
\newcounter{example}

   \newcounter{propertycounter}

\begin{document}
%
\title{Spatially Coupled Sparse Codes on \\Graphs - Theory and Practice}
%
%
%

\author{D.~J.~Costello,~Jr., L.~Dolecek, T.~E.~Fuja, J.~Kliewer, \\D.~G.~M.~Mitchell, and R.~Smarandache
\thanks{This work was supported in part by the National Science Foundation under Grant Numbers CCF-$1161754$, CCF-$1161774$, CCF-$1161798$, and CCF-$1252788$.}
\thanks{D.~J.~Costello,~Jr., T.~E.~Fuja, D.~G.~M.~Mitchell,  and R.~Smarandache are with the Department
of Electrical Engineering, University of Notre Dame, Notre Dame,
IN 46556, USA (e-mail: costello.2@nd.edu; tfuja@nd.edu; david.mitchell@nd.edu; rsmarand@nd.edu). D.~G.~M.~Mitchell and  R.~Smarandache are also with the Department of Mathematics, University of Notre Dame, Notre Dame,
IN 46556, USA.}
\thanks{L.~Dolecek is the with the Department
of Electrical Engineering, University of California, Los Angeles, Los Angeles, CA 90095, USA (e-mail: dolecek@ee.ucla.edu).}
\thanks{J.~Kliewer is with the Klipsch School of Electrical and Computer Engineering, New Mexico State University, Las Cruces, NM 88003, USA (e-mail: jkliewer@nmsu.edu).}}

%
%

\markboth{}%
{}
%



\maketitle

\begin{abstract}

Since the discovery of turbo codes 20 years ago and the subsequent re-discovery of low-density parity-check codes a few years later, the field of channel coding has experienced a number of major advances.  Up until that time, code designers were usually happy with performance that came within a few decibels of the Shannon Limit, primarily due to implementation complexity constraints, whereas the new coding techniques now allow performance within a small fraction of a decibel of capacity with modest encoding and decoding complexity.  Due to these significant improvements, coding standards in applications as varied as wireless mobile transmission, satellite TV, and deep space communication are being updated to incorporate the new techniques.  In this paper, we review a particularly exciting new class of low-density parity-check codes, called spatially-coupled codes, which promise excellent performance over a broad range of channel conditions and decoded error rate requirements. 
\end{abstract}


%
\IEEEpeerreviewmaketitle

\section{Introduction}\label{sec:intro}

\begin{figure}
\begin{center}
\includegraphics[width=4.3in]{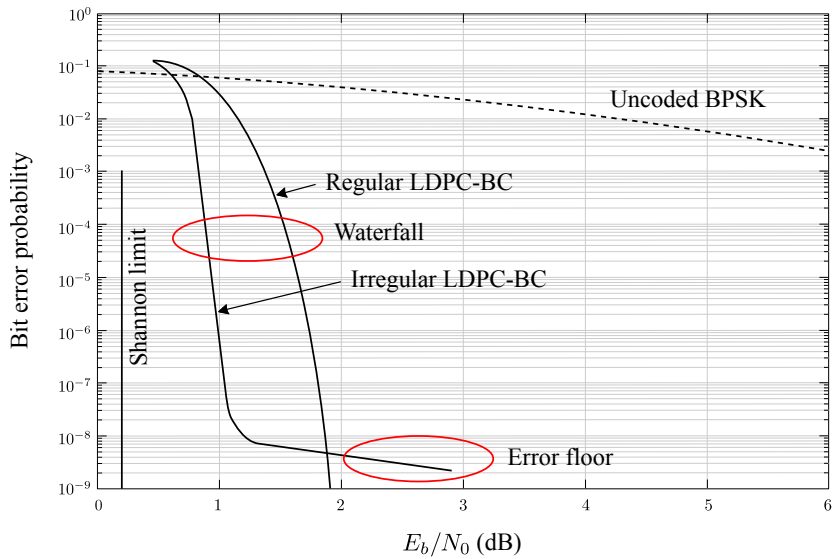}
\caption{A sketch of typical LDPC-BC decoded BER performance over the AWGNC. Also shown for comparison are the channel capacity, or Shannon limit, and the performance of uncoded binary phase-shift keying (BPSK) transmission.}\label{fig:ldpcsketch}
\end{center}
\end{figure}

%
%
%
%
\emph{Low-density parity-check (LDPC)} codes, combined with iterative \emph{belief-propagation (BP)} decoding, have emerged in recent years as the most promising method of achieving the goal set by Shannon \cite{sha48} in his landmark 1948 paper: to communicate reliably over a noisy transmission channel at a rate approaching channel capacity. Indeed, many applications have recently adopted LDPC codes as industry standards - such as wireless LANs (IEEE 802.11n), WiMax (IEEE 802.16e), digital video broadcasting (DVB-S2), 10GBase-T Ethernet (IEEE 802.3an), and the ITU-T standard for networking over power lines, phone lines, and coaxial cable (G.hn/G.9960). The key feature that sets LDPC codes apart from other capacity approaching codes is that, with suboptimal iterative BP decoding, complexity grows only linearly with code block length, resulting in practically realizable decoder implementations for powerful (long block length) codes. (The decoding complexity of optimum \emph{maximum likelihood (ML)} decoding, on the other hand, grows exponentially with block length, making it impractical for large block lengths.) \emph{LDPC block code (LDPC-BC)} designs can be classified in two types: regular and irregular. \emph{Regular} codes, as originally proposed by Gallager \cite{gal62} in 1962, are \emph{asymptotically good} in the sense that their \emph{minimum distance} grows linearly with block length. This guarantees, with ML decoding, that the codes do not suffer from the \emph{error floor} phenomenon, a flattening of the \emph{bit error rate (BER)} curve that results in poor performance at high \emph{signal-to-noise ratios (SNRs)}, and similar behavior is observed with iterative BP decoding as well. However, the iterative decoding behavior of regular codes in the so-called \emph{waterfall}, or moderate BER, region of the performance curve falls short of capacity, making them unsuitable for severely power-constrained applications, such as uplink cellular data transmission or digital satellite broadcasting systems, that must achieve the best possible performance at moderate BERs. On the other hand, \emph{irregular} codes, pioneered by Luby et al. \cite{lmss01} in 2001, exhibit capacity approaching performance in the waterfall but are normally subject to an error floor, making them undesirable in applications, such as data storage and optical communication, that require very low decoded BERs. Typical performance characteristics of regular and irregular LDPC-BCs on an \emph{additive white Gaussian noise channel (AWGNC)} are illustrated in Fig. \ref{fig:ldpcsketch}, where the channel SNR is expressed in terms of $E_b/N_0$, the \emph{information bit signal-to-noise ratio}. 

In this paper, we highlight a particularly exciting new class of LDPC codes, called \emph{spatially-coupled LDPC (SC-LDPC)} codes, which promise robustly excellent performance over a broad range of channel conditions, including both the waterfall and error floor regions of the BER curve.  We also show how SC-LDPC codes can be viewed as a type of \emph{LDPC convolutional code (LDPC-CC)}, since spatial coupling is equivalent to introducing memory into the encoding process. In channel coding parlance, the key feature of SC-LDPC codes that distinguishes them from standard LDPC codes is their ability to combine the best features of regular and irregular codes in a single design: (1) capacity approaching iterative decoding \emph{thresholds}\footnote{Roughly speaking, the threshold associated with a particular code/decoder is the lowest SNR at which the decoder can operate reliably. }, characteristic of optimized irregular codes, thus promising excellent performance in the waterfall, and (2) linear growth of minimum distance with block length, characteristic of regular codes, thus promising the elimination of an error floor. As will be discussed in more detail in Section \ref{sec:scstructure}, this is achieved by introducing a slight \emph{structured irregularity} into the Tanner graph representation of a regular LDPC code. An added feature of the SC-LDPC code design is that the resulting graph retains the essential implementation advantages associated with the structure of regular codes, compared to typical irregular designs. The research establishing the performance characteristics of SC-LDPC codes relies on ensemble average asymptotic methods, \emph{i.e.}, the capacity approaching thresholds and asymptotically good minimum distance behavior are shown to hold for typical members of SC-LDPC code ensembles\footnote{A code ensemble is the collection of all codes sharing some common set of characteristics.} as the block length tends to infinity.  (Following the lead of Shannon, coding theorists often find it easier and more insightful to analyze the average asymptotic behavior of code ensembles than to determine the exact performance of specific codes.) These research results are summarized in Section \ref{sec:scstructure}.

Section \ref{sec:scproblems} discusses issues related to realizing the exceptional promise of SC-LDPC codes with specific code and decoder designs suitable for low-complexity implementation at block lengths typically employed in practice: 1) the use of high-throughput, parallel, pipeline decoding and 2) the use of \emph{sliding-window} decoding strategies for reduced latency and computational complexity, and Section \ref{sec:op} contains a short summary of several open research problems. Finally, Section \ref{sec:conc} includes some concluding remarks along with a brief discussion of the promising use of the spatial coupling concept beyond the realm of channel coding. 

\section{SC-LDPC Codes: Basic Structure and Asymptotic Properties}\label{sec:scstructure}
\begin{figure}
\begin{center}
\raisebox{14mm}{$\mathbf{H} = $}\hspace{2mm}\includegraphics[width=5in]{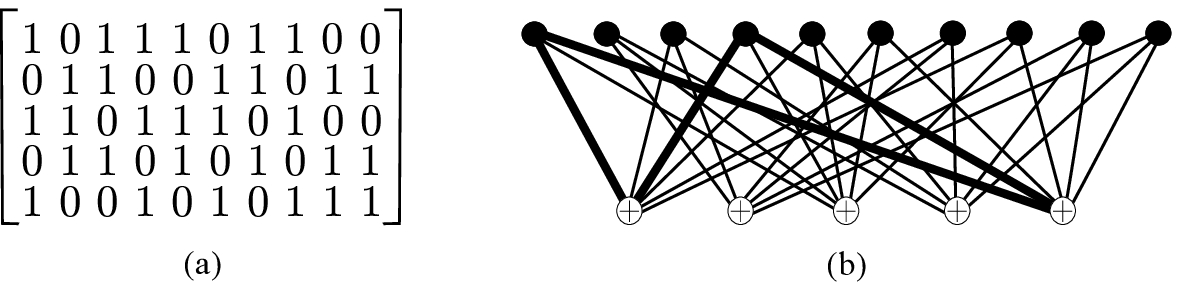}
\caption[the caption]{(a) Parity-check matrix of a $(3,6)$-regular LDPC-BC with block length $n=10$ and (b) the associated $(3,6)$-regular Tanner graph. The filled circles \raisebox{-0.5mm}{\includegraphics[width=2.25mm]{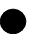}} represent code bits, or variable nodes,  the open circles \raisebox{-0.5mm}{\includegraphics[width=2.25mm]{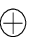}} represent parity-checks, or constraint nodes, and the darkened edges represent a cycle of length $4$.}\label{fig:tannergraph}
\end{center}
\end{figure}

A $(J,K)$\emph{-regular} LDPC-BC of \emph{rate} $R=k/n$ and \emph{block length} $n$ is defined as the null space of an $(n-k)\times n$  binary parity-check matrix $\mathbf{H}$, where each row of $\mathbf{H}$ contains exactly $K$ ones, each column of $\mathbf{H}$ contains exactly $J$ ones, and both $J$ and $K$ are small compared with the number of rows in $\mathbf{H}$. An LDPC code is called \emph{irregular} if the row and column weights are not constant. It is often useful to represent the parity-check matrix $\mathbf{H}$ using a bipartite graph called the \emph{Tanner graph}. In the Tanner graph representation, each column of $\mathbf{H}$ corresponds to a \emph{code bit} or \emph{variable node} and each row corresponds to a \emph{parity-check} or \emph{constraint node}. If position $(i,j)$ of $\mathbf{H}$ is equal to one, then constraint node $i$ is connected by an \emph{edge} to variable node $j$ in the Tanner graph; otherwise, there is no edge connecting these nodes. Fig. \ref{fig:tannergraph} depicts the parity-check matrix and associated Tanner graph of a $(3,6)$-regular LDPC-BC. In this example, we see that all variable nodes have \emph{degree} $3$, since they are connected to exactly $3$ constraint nodes, and similarly all constraint nodes have degree $6$. In the case of irregular codes, the notion of \emph{degree distribution} is used to characterize the variations of constraint and variable node degrees (see \cite{lmss01}).

Using the Tanner graph, iterative BP decoding can be viewed as passing messages back and forth between variable and constraint nodes (see, \emph{e.g.}, \cite{ru08}). On an AWGNC, for example, the messages are typically \emph{log-likelihood ratios (LLRs)} associated with the (in general soft-valued) received symbols, which serve as indicators of the probability that a particular code bit is a ``1'' or a ``0''. These LLRs are then passed across the graph and adjusted iteratively to reflect the parity constraints until some stopping condition is satisfied, indicating that the received symbols can be reliably decoded.

Certain properties of the Tanner graph can serve as useful indicators of the performance characteristics of iterative decoding. In Fig. \ref{fig:tannergraph}(b), the darkened edges indicate a \emph{cycle} of length $4$. In general, codes with short cycles do not perform well in the waterfall, due to the build up of correlation in the iterative process. Hence it is desirable to choose codes with large \emph{girth}, the length of the shortest cycle, for good waterfall performance. In terms of the error floor performance, minimum distance is the best indicator for ML decoding, and asymptotically good codes are typically not subject to an error floor. For iterative decoding, however, certain substructures of the Tanner graph, such as \emph{trapping sets} and \emph{absorbing sets}\footnote{An absorbing set  is a subset of variable nodes and constraint nodes (a subgraph) that can cause an iterative decoder to get stuck (fail to decode) if the variable nodes in the subset receive unreliable LLRs from the channel.  In other words, the decoder may be unable to \emph{correct errors} associated with the variable nodes in an absorbing set.}  (an important subclass of trapping sets), can cause iterative decoders to fail apart from minimum distance considerations, resulting in the emergence of an error floor. Hence it is desirable to select graphs without problematic trapping or absorbing sets for good error floor performance.

SC-LDPC codes can be viewed as a type of LDPC-CC, first introduced in the open literature by Jimenez-Felstrom and Zigangirov in 1999 \cite{fz99}.  A rate $R = b/c$ LDPC-CC can be represented by a bi-infinite parity-check matrix
\begin{equation}\label{Hmat}
\mathbf{H}_{cc}=\left[\begin{array}{llllllll}
\multicolumn{1}{c}{\ddots}&\multicolumn{1}{c}{\ddots} &&&\multicolumn{1}{c}{\ddots}\\
 \mathbf{H}_{m_s}(t-1)\hspace{-1mm} & \mathbf{H}_{m_s-1}(t-1)\hspace{-1mm} &  \multicolumn{1}{r}{\cdots}&&\mathbf{H}_{0}(t-1)\hspace{-1mm}\\
& \mathbf{H}_{m_s}(t) & \mathbf{H}_{m_s-1}(t) &  \multicolumn{1}{r}{\cdots}&& \mathbf{H}_{0}(t)\\
&& \mathbf{H}_{m_s}(t+1)\hspace{-1mm} & \mathbf{H}_{m_s-1}(t+1)\hspace{-1mm} &  &\cdots& \mathbf{H}_{0}(t+1)\\
&& \multicolumn{1}{c}{\ddots} & \multicolumn{1}{c}{\ddots}&&  & \multicolumn{1}{c}{\ddots}\\
                           \end{array}\right],\end{equation}
composed of a diagonal band of $(c - b) \times c$ submatrices $\mathbf{H}_i(t)$, $0 \leq i \leq m_s$, $t = 0,1,2,\ldots$, where the rows and columns of $\mathbf{H}_{\mathrm{cc}}$ are sparse, \emph{i.e.}, they contain a small number of non-zero entries. If $\mathbf{H}_{\mathrm{cc}}$ contains only zeros and ones, the code is binary; otherwise, it is non-binary.  $m_s$ is called the \emph{syndrome former memory}, where $m_s + 1$ is the width of each row in submatrices, and $\nu_s = (m_s + 1)c$, the width of each row in symbols, is called the \emph{decoding constraint length}. If $\mathbf{H}_{\mathrm{cc}}$ contains a fixed number $J$ of ones in each column and a fixed number $K$ of ones in each row, it represents a $(J, K)$-regular LDPC-CC; otherwise, the code is irregular. In general, $\mathbf{H}_{\mathrm{cc}}$ describes a \emph{time-varying} LDPC-CC, and if the rows of $\mathbf{H}_{\mathrm{cc}}$ vary periodically, the code is \emph{periodically time-varying}. If the rows of $\mathbf{H}_{\mathrm{cc}}$ do not vary with time, the code is \emph{time-invariant}.

\begin{figure}[t]
\begin{center}
\begin{picture}(480,341)
\put(30,0){\includegraphics[width=5in]{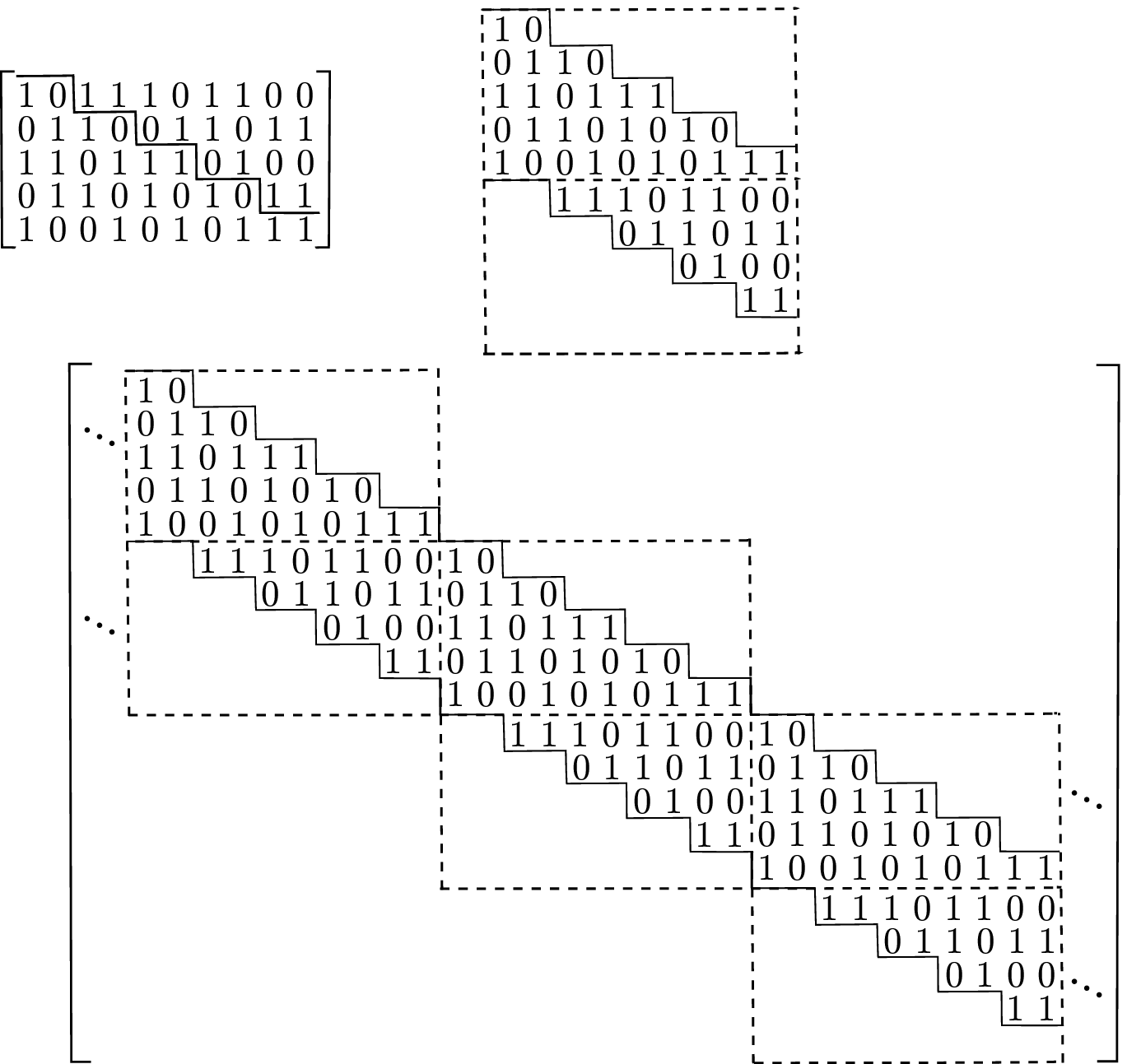}}
\put(0,111){$\leadsto\mathbf{H}_{\mathrm{cc}}=$}
\put(0,288){$\mathbf{H}=$}
\put(155,288){$\leadsto$}
\put(400,118){(diagonal matrix}
\put(400,105){extension)}
\put(300,288){(cut-and-paste)}
\end{picture}\\\vspace{2mm}
\includegraphics[width=5.8in]{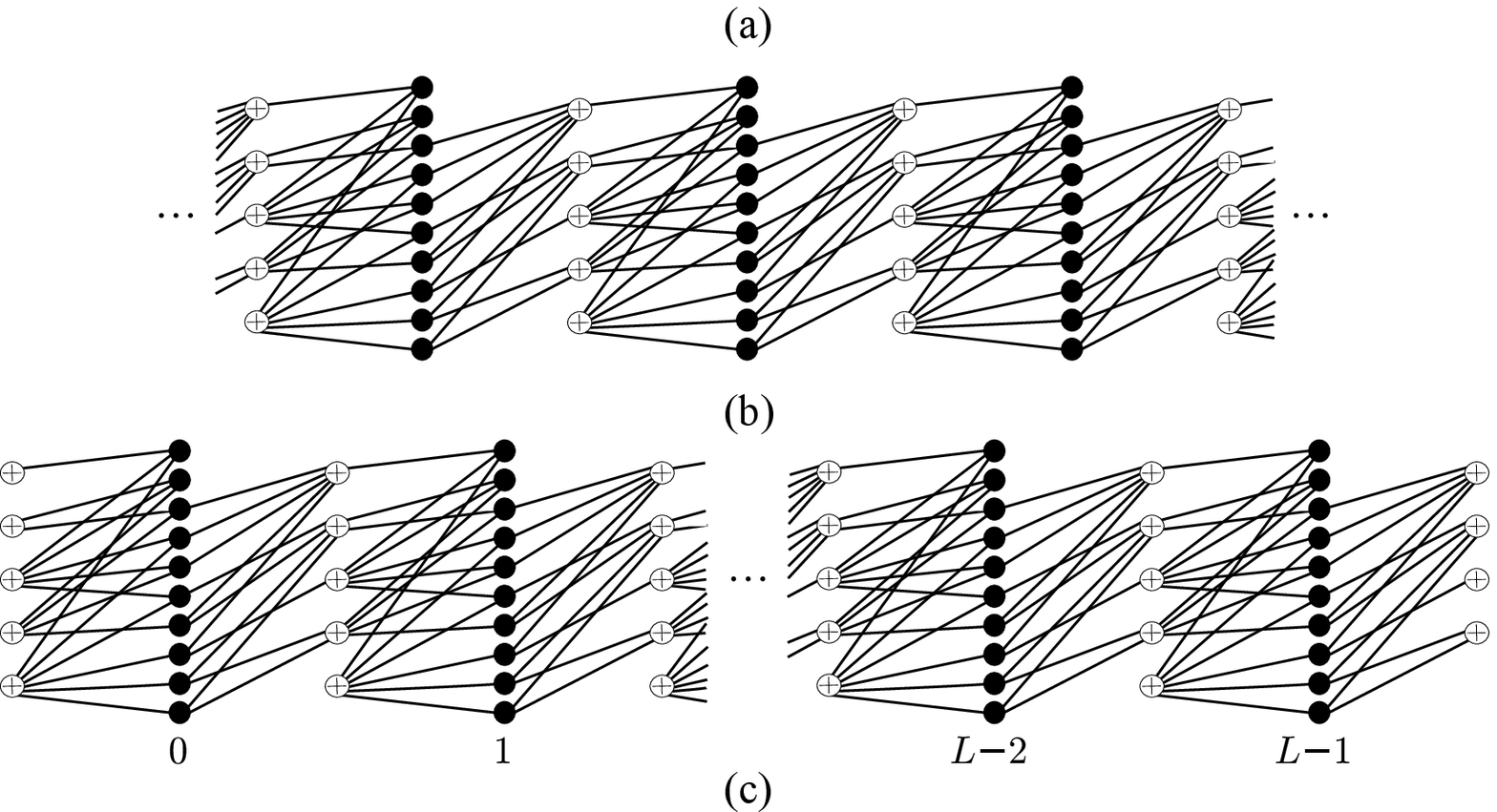}
\caption{(a) An illustration of the unwrapping procedure for a $(3,6)$-regular LDPC-BC, (b) the Tanner graph associated with the unwrapped $(3,6)$-regular LDPC-CC, and (c) the terminated Tanner graph associated with the unwrapped $(3,6)$-regular LDPC-CC.}\label{fig:unwrapping}
\end{center}\vspace{-5mm}
\end{figure}

Using a technique termed \emph{unwrapping} in \cite{fz99}, it is possible to take any good LDPC-BC and \emph{unwrap} it to form an LDPC-CC with improved BER performance.  The unwrapping procedure applies cut-and-paste and diagonal matrix extension operations to the parity-check matrix $\mathbf{H}$ of an LDPC-BC to produce a bi-infinite parity-check matrix $\mathbf{H}_{\mathrm{cc}}$ of an LDPC-CC, as illustrated in Fig.~\ref{fig:unwrapping}(a), where $\mathbf{H}$ represents a $(3,6)$-regular block code with block length $n=10$ and  $\mathbf{H}_{\mathrm{cc}}$ represents a $(3,6)$-regular convolutional code with constraint length $\nu_s = 10$. The bi-infinite (convolutional) Tanner graph representation of $\mathbf{H}_{\mathrm{cc}}$ is shown in Fig.~\ref{fig:unwrapping}(b), and we see that the unwrapping procedure preserves the graph structure of the underlying LDPC-BC, \emph{i.e.}, all node degrees remain the same and the local connectivity of nodes is unchanged.

\begin{figure}[h]
\begin{center}
\includegraphics[width=4in]{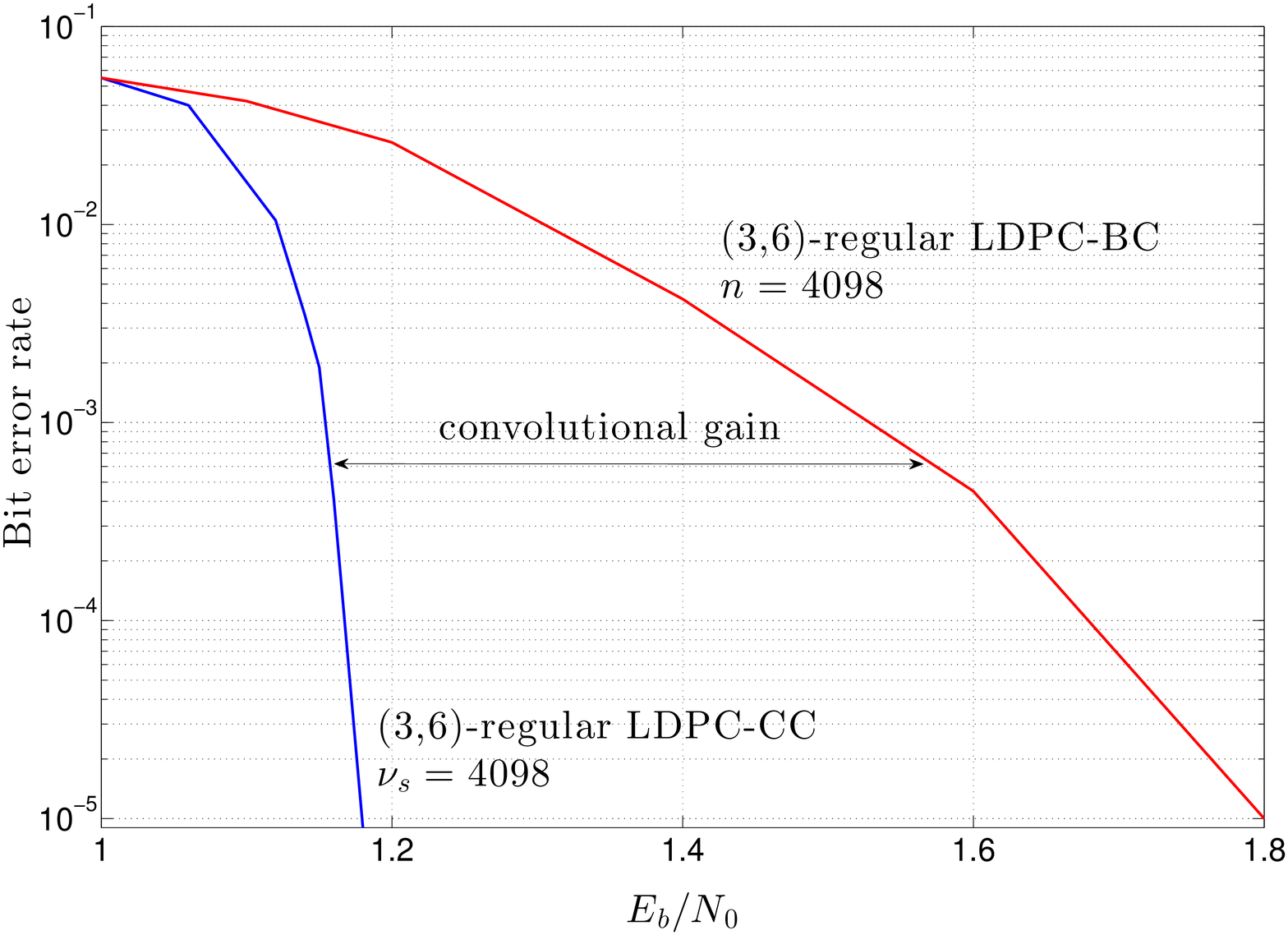}
\caption{An illustration of convolutional gain.}\label{fig:convgain}
\end{center}
\end{figure}

Extensive computer simulation results (see, \emph{e.g.}, \cite{psvc11}) have verified that, for practical code lengths, LDPC-CCs obtained by unwrapping an LDPC-BC achieve a substantial \emph{convolutional gain} compared to the underlying LDPC-BC, where both codes have the same computational complexity with iterative decoding and the block length of the LDPC-BC equals the constraint length of the LDPC-CC.  An example illustrating this convolutional gain is shown in Fig.~\ref{fig:convgain}.

Even though the Tanner graph representation of an LDPC-CC extends infinitely both forward and backward in time, in practice there is always some finite starting and ending time, \emph{i.e.}, the Tanner graph is \emph{terminated} at both the beginning and the end (see Fig. \ref{fig:unwrapping}(c)). A remarkable feature of this graph termination, first noted numerically in the paper by Lentmaier et al. \cite{lscz10} for both the \emph{binary erasure channel (BEC)} and the AWGNC and then shown analytically (for the BEC) by Kudekar et al. \cite{kru11}, is the so-called \emph{threshold saturation} effect. Consider for purposes of illustration the $(3, 6)$-regular LDPC-BC ensemble with AWGNC iterative BP decoding threshold $E_b/N_0 =1.11$dB, which is also the threshold of the associated (unterminated) LDPC-CC ensemble. As the graph termination length $L$ becomes large, the threshold of the (terminated) LDPC-CC ensemble improves all the way to $0.46$dB, the threshold of the $(3, 6)$-regular LDPC-BC ensemble with ML decoding.\footnote{For small termination lengths $L$, the terminated LDPC-CC suffers a rate loss compared to the underlying LDPC-BC, but this rate loss vanishes for large $L$.}  In other words, terminated LDPC-CCs with BP decoding are capable of achieving the same performance as comparable LDPC-BCs with (much more complex, and impractical) ML decoding! This ``step-up'' of the BP threshold to the ML threshold is referred to as threshold saturation. Note that, after termination, the LDPC-CC code ensemble can be viewed as an LDPC-BC ensemble with block length $n=(m_s+1)cL=\nu_s L$. However, compared to typical LDPC-BC designs that have no restrictions on the location of the ones in the parity-check matrix and hence allow connections across the entire graph, the LDPC-CC code ensemble has a highly \emph{localized} graph structure, since the non-zero portion of the parity-check matrix is restricted to a diagonal band of width $\nu_s$. We will see later that this structure, in addition to yielding excellent iterative decoding thresholds, also gives rise to an efficient decoder implementation.

Threshold saturation is a result of the termination, which introduces a slight structured irregularity in the graph. Termination has the effect of introducing lower constraint node degrees, \emph{i.e.}, a structured irregularity, at each end of the graph (see Fig.~\ref{fig:unwrapping}(c)). In the context of iterative BP decoding, the smaller degree constraint nodes pass more reliable messages to their neighboring variable nodes, and this effect propagates throughout the graph as iterations increase. This results in BP thresholds for terminated LDPC-CC ensembles that, for large enough degree densities ($J$ and $K$ for regular codes), actually \emph{achieve capacity} as the constraint length $\nu_s$ and the termination length $L$ go to infinity. In addition, for regular LDPC-CCs, the terminated (slightly irregular) ensembles are still asymptotically good, in the sense that their minimum distance grows linearly with block length $n$.

\begin{figure}[h]
\begin{center}
\includegraphics[width=\columnwidth]{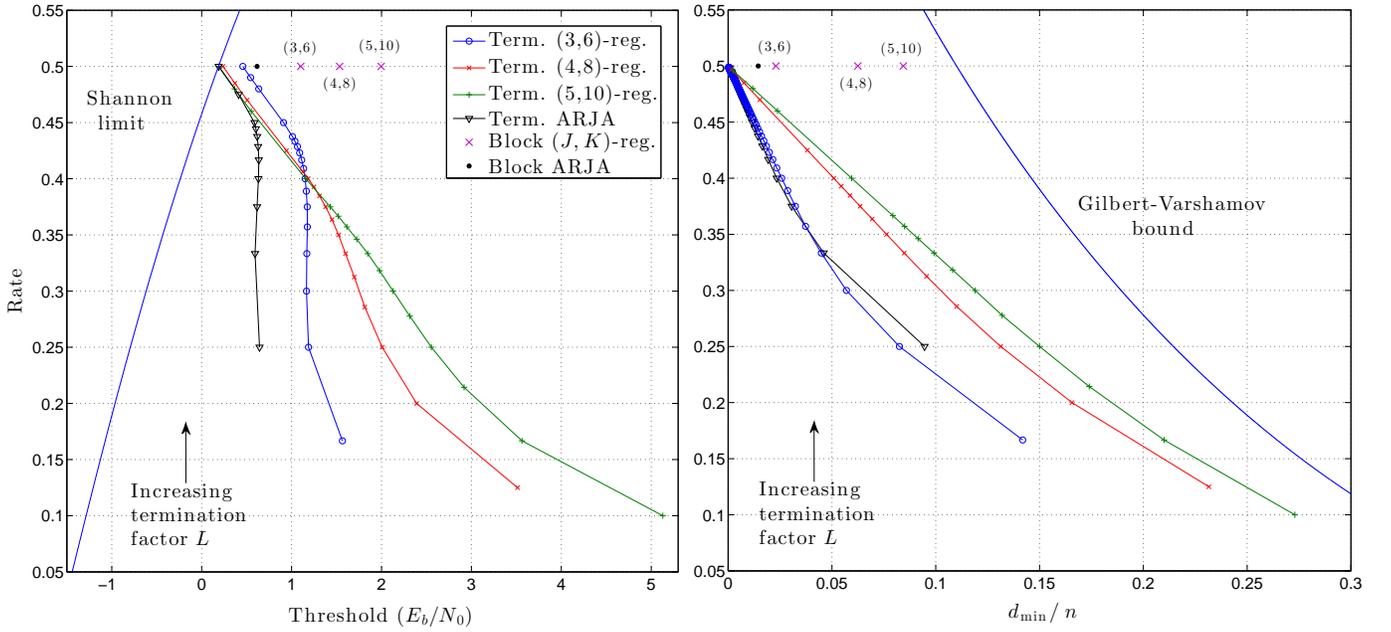}
\caption{AWGNC BP decoding thresholds (left) and minimum distance growth rates (right) for families of terminated $(J, 2J)$-regular LDPC-CC ensembles, terminated ARJA-based LDPC-CC ensembles, and the underlying LDPC-BC ensembles. The Shannon limit (lower bound on $E_b/N_0$) and the Gilbert-Varshamov bound (upper limit on $d_{\mathrm{min}}/n$) are plotted for comparison.}\label{fig:tradeoff}
\end{center}
\end{figure}

The net result of these effects is captured in Fig.~\ref{fig:tradeoff}, which illustrates the tradeoffs between the AWGNC BP decoding threshold (in $E_b/N_0$), the minimum distance growth rate ($d_\mathrm{min}/n$), and the code rate ($R$) for several $(J, 2J)$-regular terminated LDPC-CC ensembles as a function of the termination length $L$. We observe that, in general, as the termination length $L$ increases, the LDPC-CC rate approaches the rate of the underlying LDPC-BC and the BP thresholds of the terminated LDPC-CC ensembles approach capacity as $J$ increases.\footnote{The BP thresholds of the terminated LDPC-CC ensembles are approaching the ML thresholds of the underlying LDPC-BC ensembles, which in turn approach capacity as $J$ increases.}  Also, linear distance growth is maintained for any finite value of $L$. 
In addition to regular ensembles, Fig.~\ref{fig:tradeoff} also includes terminated LDPC-CC ensembles based on the irregular ARJA codes designed by Divsalar et al. \cite{ddja09}, an irregular LDPC-BC ensemble with linear distance growth and better thresholds than comparable regular ensembles. (Irregular LDPC-BC ensembles with optimized degree profiles already have thresholds close to capacity, and they do not possess linear distance growth, so little is to be gained by applying the terminated LDPC-CC construction in these cases.) The major advantage of the regular terminated LDPC-CC constructions highlighted above is that they can achieve the same thresholds as the optimized irregular designs \emph{without sacrificing} linear distance growth, while maintaining the desirable structural features of regular codes \cite{lmfc10b}.

\begin{figure}[h]
\begin{center}
\includegraphics[width=\columnwidth]{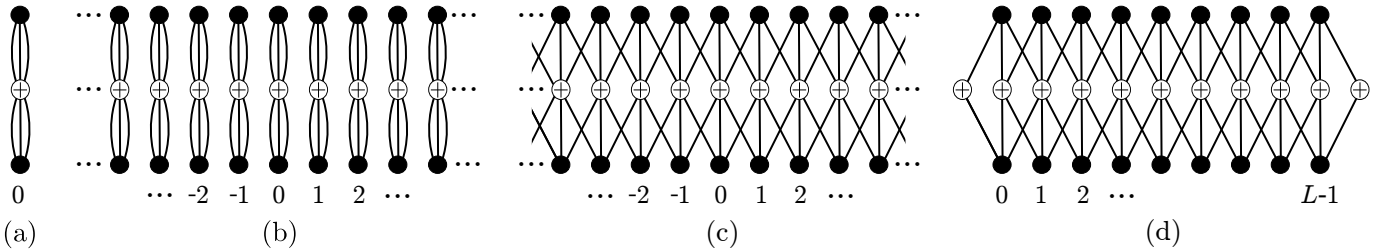}
\caption{Tanner graphs associated with (a) a $(3, 6)$-regular block code protograph, (b) a bi-infinite chain of uncoupled $(3, 6)$-regular protographs, (c) a bi-infinite chain of coupled $(3, 6)$-regular protographs, and (d) a terminated chain of coupled $(3, 6)$-regular protographs. 
}\label{fig:coupling}
\end{center}
\end{figure}

An insightful way of viewing the design of terminated LDPC-CCs is to use a \emph{protograph} representation of the code ensemble \cite{tho03}. A block code protograph is a small bipartite graph, with $c$ variable nodes and $c - b$ constraint nodes, that is used to represent the parity-check matrix of a  rate $R=b/c$ block code with block length $c$, where $c$ and $b$ are typically small integers.  An example of a  block code protograph with $c = 2$ variable nodes of degree $3$ and $c - b = 1$ constraint node of degree $6$ is shown in Fig.~\ref{fig:coupling}(a).  The corresponding $(c - b) \times c = 1\times 2$ parity-check matrix in this case is given by $\mathbf{B} = [\>3, 3\>]$, where  $\mathbf{B}$ is called the \emph{base} matrix and the entries in $\mathbf{B}$ denote that the constraint node in the graph is connected by $3$ edges to each  of the two variable nodes. Now form an $M$-fold \emph{graph cover} by applying the \emph{graph lifting} operation \cite{psvc11} to the protograph. This can be represented by placing a randomly selected permutation of size $M$ (typically a large integer) on each edge of the protograph, thereby connecting $M$ copies of each node and resulting in the Tanner graph of a $(3, 6)$-regular LDPC-BC with rate $R=b/c$ and block length $n=Mc$.  The collection of all $M$-fold graph covers then represents a $(3, 6)$-regular LDPC-BC ensemble. The graph lifting operation is equivalent to replacing each entry in the base matrix $\mathbf{B}$ with the (modulo-$2$) sum of $3$ randomly chosen (and distinct) $M \times M$ permutation matrices.\footnote{Depending on the permutations selected, the lifted parity-check matrix may contain some redundant rows, resulting in a code rate slightly larger than $b/c$.}

In Fig.~\ref{fig:coupling}(b), a bi-infinite chain of uncoupled block code protographs is shown, corresponding to block code transmission over time. Now consider \emph{spreading the edges} of each protograph so that they connect to one or more nearby protographs in the chain in such a way that the degrees of all the nodes are preserved, as illustrated in Fig.~\ref{fig:coupling}(c). This has the effect of \emph{coupling} the different protographs together in a bi-infinite chain, which is equivalent to introducing \emph{memory} into the code design, \emph{i.e.}, transitioning from a rate $R = 1/2$ block code to a rate $R = 1/2$ convolutional code, where the syndrome former memory $m_s$, the coupling depth or number of nearby protographs connected to a given protograph, equals $2$ in this case.  Viewing (\ref{Hmat}) as the base matrix $\mathbf{B}_{\mathrm{cc}}$ of the coupled \emph{convolutional protograph}, this \emph{edge-spreading} (unwrapping) technique is equivalent to splitting the block code base matrix $\mathbf{B}$ into $m_s+1$ component submatrices $\mathbf{B}_0, \mathbf{B}_1, \ldots, \mathbf{B}_{m_s}$, such that $\mathbf{B} = \mathbf{B}_0 + \mathbf{B}_1 + \cdots + \mathbf{B}_{m_s}$, where in the most general time-varying case each row of (\ref{Hmat}) could start with a different base matrix (\emph{i.e.}, each protograph in the chain could be different) and the edge-spreading associated with each protograph could also vary.  For the case shown in Fig.~\ref{fig:coupling}(c), the coupled convolutional protograph is time-invariant and the edge spreading is given by
$\mathbf{B} = [\>3,3\>] = \mathbf{B}_0 + \mathbf{B}_1 + \mathbf{B}_2 = [\>1, 1\>] + [\>1, 1\>] + [\>1, 1\>],$ so the bi-infinite convolutional base matrix becomes
\begin{equation}
 \mathbf{B}_{\mathrm{cc}}=\left[
\begin{array}{cccccccccc}
  \ddots\hspace{-2mm} &   & \ddots\hspace{-2mm} &   & \ddots\hspace{-1.8mm} & & & & &\\
1 & 1 & 1 & 1 & 1 & 1 & & & &\\
 & & 1 & 1 & 1 & 1 & 1 & 1 & &\\
& & & & 1 & 1 & 1 & 1 & 1 & 1\\
& & & & & \ddots\hspace{-2mm} & & \ddots\hspace{-2mm} & & \ddots\hspace{-2mm}
\end{array}
\right].
\end{equation}
If the graph lifting operation is now applied to the convolutional protograph by placing randomly selected permutations of size $M$ on each edge of the graph, an unterminated $(3, 6)$-regular LDPC-CC ensemble with constraint length $\nu_s = (m_s+1)Mc$ results.

The coupled convolutional protograph can then be terminated, resulting in reduced constraint node degrees at both ends, as shown in Fig.~\ref{fig:coupling}(d), and the $(L+2)\times 2L$ terminated convolutional base matrix becomes

\begin{equation}
 \mathbf{B}_{\mathrm{cc}}=\left[
\begin{array}{ccccccc}
1 & 1 & &  &  &  &  \\\vspace{-1.7mm}
 1&1 & 1 & 1 &  &  &    \\\vspace{-1.7mm}
1& 1& 1& 1& \ddots &  &     \\\vspace{-1.7mm}
& & 1& 1& \ddots&     1  &1 \\
& & & & \ddots&      1 &1 \\
& & & & &      1 &1 \\
\end{array}
\right].
\end{equation}
Now applying the graph lifting operation results in a terminated $(3, 6)$-regular LDPC-CC ensemble, which can also be viewed as an LDPC-BC with block length $n=LMc$.  Note that, because of the reduced constraint node degrees at each end, the graph is not quite $(3, 6)$-regular, and the code rate associated with the terminated LDPC-CC ensemble is less than the rate $R = 1/2$ of the underlying LDPC-BC.  However, as the termination length $L \to \infty$, the terminated LDPC-CC ensemble becomes $(3, 6)$-regular and the associated code rate $R \to 1/2$.  

Because the memory employed in the convolutional code design has the effect of coupling together several identical block code protographs, the above graphical construction of terminated LDPC-CC ensembles, also denoted as SC-LDPC code ensembles, is referred to as \emph{spatial coupling} \cite{kru11}. While it is the asymptotic threshold and minimum distance properties of SC-LDPC code ensembles, summarized above, that have generated so much interest in these codes, some basic questions having to do with how best to employ them for practical code lengths still must be solved before they can realize their exceptional promise as a robust, near-optimal solution to the channel coding problem. The following section describes some of these practical issues.

\section{SC-LDPC Codes: Decoding and Practical Considerations}\label{sec:scproblems}
\begin{figure}[h]
\begin{center}
\includegraphics[width=7in]{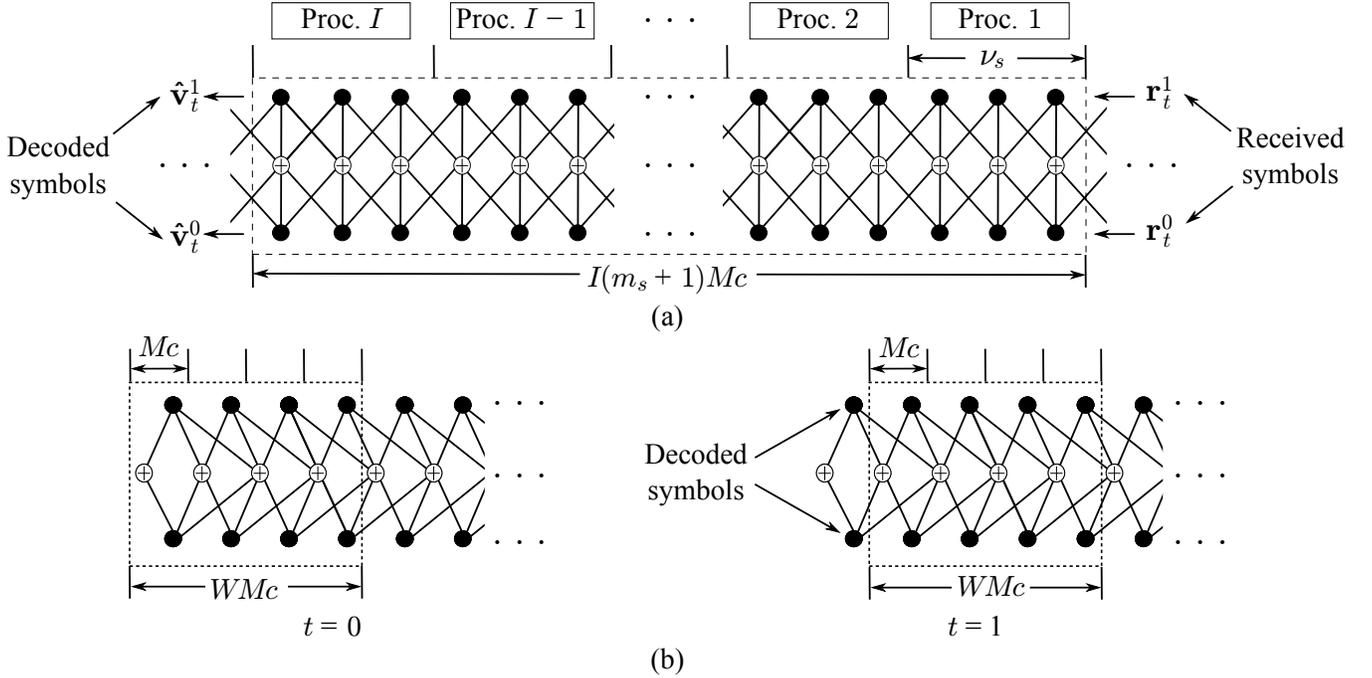}
\end{center}
\caption{(a) Example of a pipeline decoder operating on the protograph of a $(3, 6)$-regular $R = 1/2$ LDPC-CC. (b) Example of a sliding window decoder with window size $W = 4$ operating on the protograph of a $(3, 6)$-regular SC-LDPC code at times $t = 0$ (left), and $t = 1$ (right).}\label{fig:tanner}
\end{figure}

An important contribution of \cite{fz99} was the introduction of a parallel, high-speed, pipeline-decoding architecture for LDPC-CCs based on the same iterative BP decoding algorithm used to decode LDPC-BCs. This is illustrated in terms of the convolutional protograph associated with an example $(3, 6)$-regular rate $R = b/c = 1/2$ LDPC-CC with $m_s = 2$ and $\nu_s=(m_s+1)Mc=6M$ in Fig.~\ref{fig:tanner}(a).  Given some fixed number $I$ of decoding iterations, the pipeline decoding architecture employs $I$ identical copies of a message-passing processor operating in parallel. Each processor covers a span of $\nu_s$ variable nodes, so that during a single decoding iteration messages are always passed within a single processor. As each new set of $Mc$ (in general soft-valued) symbols (represented by $M$-ary vectors $\mathbf{r}_t^0$ and $\mathbf{r}_t^1$ in Fig. \ref{fig:tanner}(a)) enters the decoder from the channel, $Mc$ new LLRs are computed and each processor updates (in parallel) exactly one set of $Mc$ variable nodes and one set of $M(c - b)$ constraint nodes.\footnote{Note that, within each processor, it is possible to trade between high-speed (parallel) operation and lower-speed (serial) operation by varying the relation between code memory $m_s$ and graph lifting factor $M$. For example, for a given constraint length $\nu_s$ (which determines the code strength), large $M$ and small $m_s$ result in high-speed processing, whereas the processing is slower for small $M$ and large $m_s$.} When the next set of $Mc$ symbols arrives, the decoding window, containing $I\nu_s$ variable nodes and $I$ processors, shifts by one time unit (corresponding to a set of $Mc$ received symbols) to the right and another decoding iteration is performed. In this fashion, the decoder continuously accepts $Mc$ new symbols from the channel and produces (with a delay of $I(m_s + 1)$ time units, or $I(m_s+1)Mc=I\nu_s$ received symbols) decoding estimates of $Mc$ symbols (represented by $M$-ary vectors $\hat{\mathbf{v}}_t^0$ and $\hat{\mathbf{v}}_t^1$ in Fig. \ref{fig:tanner}(a)) at each time unit. 

As noted above, for the parallel pipeline decoder architecture illustrated in Fig.~\ref{fig:tanner}(a), the hardware processor includes only one constraint length, \emph{i.e.}, $\nu_s = (m_s + 1)Mc$ variable nodes. On the other hand, in the case of SC-LDPC codes (terminated LDPC-CCs), the standard LDPC-BC decoder architecture includes all the variables nodes in a block, which equals $LMc$, \emph{i.e.}, the total block length (see Fig.~\ref{fig:coupling}(d)). Since, typically, $L >> m_s$, the pipeline architecture achieves a large saving in processor size compared to the standard LDPC-BC architecture, while, for the same number of iterations, the performance of the two decoders is identical. 

The latency and memory requirements of the pipeline architecture, however, involve an additional factor of the number of iterations $I,$ so that $I\nu_s = I(m_s + 1)Mc$ represents the total decoding latency in received symbols and the total number of soft received values that must be stored in the decoder memory at any given time. This equals the length of the decoding window in Fig.~\ref{fig:tanner}(a), where the factor of $M$ is included to account for the size of the permutation matrix. In some applications, since capacity-approaching performance can require a large number of iterations, these latency and storage requirements may be unacceptably high. 

This fact has spurred interest in a modified \emph{sliding window} decoding architecture for SC-LPDC codes with much reduced latency and memory requirements \cite{ips+12}. Rather than maintaining a full window size of $I\nu_s = I(m_s + 1)Mc$ symbols, like the pipeline decoder, the sliding window decoder uses a much smaller window, typically just a few constraint lengths. The concept of a sliding window decoder is illustrated in Fig.~\ref{fig:tanner}(b). Assuming a window size of $WMc$ symbols, where $W$ is the number of protograph sections within the window, decoding iterations proceed until some stopping criterion is met or a fixed number of iterations has been performed, after which the window shifts and the $Mc$ symbols shifted out of the window are decoded. The key feature of a sliding window decoder is that, for $W << I(m_s + 1)$, its latency and memory requirements are much less than for the pipeline decoder. Values of $W$ roughly $5$ to $10$ times as large as $(m_s+1)$ have been shown (see \cite{lpf11}) to result in significant savings in latency and memory with minimal performance degradation, compared to using the full window size $I(m_s+1)$ for the (typically large) fixed number of iterations $I$ needed to optimize the performance of the pipeline decoder.\footnote{Even for a sliding window decoder extending over a small number of constraint lengths, it may still be advantageous to employ the parallel pipeline architecture in the implementation of the decoding window.} 

Because the initial few (depending on the coupling depth, or code memory) positions of the window include a part of the graph with reduced constraint node degrees (see Fig.~\ref{fig:tanner}(b)), the information passed to variable nodes during the iterations associated with these initial window positions is highly reliable. The design of the sliding window decoder insures that this highly reliable information then propagates down the graph as the window is shifted. This phenomenon is responsible for the threshold saturation effect associated with SC-LDPC codes. The same phenomenon manifests itself with the standard LDPC-BC or pipeline decoding architectures, but recent work shows that the propagation of reliable information through the graph occurs more efficiently with the sliding window architecture \cite{lpf11}. Thus, besides reducing latency and memory, another motivation for considering a sliding window decoder is to reduce the number of node updates (computational complexity) required to achieve a given level of performance. 

\section{Open Research Problems}\label{sec:op}
Below is a partial list of open research problems related to the practical realization of SC-LDPC codes.
\begin{itemize}
\item A topic of signficant current research interest involves a detailed performance/complexity comparison of SC-LDPC codes with LDPC-BCs. In particular, can SC-LDPCs with sliding window decoding achieve better performance than standard LDPC-BCs with less computational complexity? A fair comparison must consider decoders with the same latency and memory requirements, \emph{i.e.}, the sliding window decoder for an SC-LDPC code must be compared to an LDPC-BC whose block length is equal to $WMc$ rather than $LMc$.  In addition, the degree profiles of the two codes should be the same, \emph{e.g.}, a $(J, K)$-regular SC-LDPC code should be compared to a $(J, K)$-regular LDPC-BC. Some of the factors to be considered include choosing the most efficient (message passing) \emph{node update schedules} for decoding and designing appropriate \emph{stopping rules} for deciding when enough iterations have been performed to allow reliable decoding. For example, practical stopping rules for the sliding window decoder  could be based on a partial syndrome check, analogous to the stopping rule normally applied to the decoding of LDPC-BCs, or on the LLR statistics associated with the next set of symbols to be decoded, for example by using a threshold criterion.

\item The results of \cite{lpf11} imply tradeoffs favorable to SC-LDPC codes (assuming sliding window decoding) compared to LDPC-BCs in the waterfall region of the BER curve. There has been only limited investigation, however, regarding the error floors that might result from the use of SC-LDPC codes. Since error floor performance is a major factor in selecting codes for applications that require very low BERs, such as data storage and optical communication, it is important to consider those factors that contribute to decoding failures for SC-LDPC codes in the error floor. An important aspect of such an analysis involves establishing the precise connection between the problematic graphical substructures (trapping sets, absorbing sets) that can cause decoding failures and various decoding parameters, including scheduling choices (parallel or serial node updates, for example), amount of quantization for the stored LLRs, and window size (for sliding window decoding).

\item SC-LDPC codes are known to have excellent asymptotic properties, but there are many open questions regarding code design, in particular for short-to-moderate block lengths: scaling the lifting factor $M$ and the termination length $L$ to achieve the best possible performance for SC-LDPC codes of finite length; studying puncturing as a means of obtaining SC-LDPC codes with high rates, so as to provide rate flexibility for standards applications; finding ways to mitigate the rate loss associated with short-to-moderate block length SC-LDPC codes, such as puncturing and partial termination, without affecting performance; and exploiting the potential of connecting together multiple SC-LDPC code chains. In addition, analyzing and designing more powerful SC-LDPC codes, such as non-binary or generalized SC-LDPC codes, may be attractive in areas such as coded modulation and flash memories.
\item Members of an LDPC code ensemble that are \emph{quasi-cyclic} (QC) are of particular interest to code designers, since they can be encoded with low complexity using simple feedback shift-registers and their structure leads to efficiencies in decoder design. The practical advantages of QC-LDPC-BC designs also carry over to the design of QC-SC-LDPC codes. It should be noted, however, that once the QC constraint is applied, the asymptotic ensemble average properties noted in Sec. \ref{sec:scstructure} do not necessarily hold (since we are now choosing a code from a restricted sub-ensemble), and particular QC-SC-LDPC codes must be carefully designed to insure good performance.\end{itemize}

\section{Concluding Remarks}\label{sec:conc}
In this paper we have attempted to provide a brief tutorial overview of the exciting new field of spatially coupled low-density parity-check codes.  Capacity approaching iterative decoding thresholds and asymptotically good minimum distance properties make these codes potentially very attractive for future industry standard applications.  We traced the origins of SC-LDPC codes to the development of low-density parity-check convolutional codes in \cite{fz99}, we used a visually convenient protograph representation to describe their construction, we discussed several issues related to practical decoder implementations, and we summarized a few still remaining open research problems. 

Finally, it has recently been shown that the improved thresholds associated with spatial coupling apply generally regardless of the particular physical situation or communication channel (see, \emph{e.g.}, \cite{kru12,kymp12}). Further, although space limitations do not allow us to provide details, we note that the concept of spatial coupling of a sequence of identical copies of a small structured graph (a protograph) is applicable to a wide variety of problems and has been shown to lead to improved system performance in areas as diverse as multiterminal source and channel coding, cooperative relaying, compressed sensing, secure communication, and statistical physics. 

\begin{thebibliography}{10}
\baselineskip 12pt
\providecommand{\url}[1]{#1}
\csname url@samestyle\endcsname
\providecommand{\newblock}{\relax}
\providecommand{\bibinfo}[2]{#2}
\providecommand{\BIBentrySTDinterwordspacing}{\spaceskip=0pt\relax}
\providecommand{\BIBentryALTinterwordstretchfactor}{4}
\providecommand{\BIBentryALTinterwordspacing}{\spaceskip=\fontdimen2\font plus
\BIBentryALTinterwordstretchfactor\fontdimen3\font minus
  \fontdimen4\font\relax}
\providecommand{\BIBforeignlanguage}[2]{{%
\expandafter\ifx\csname l@#1\endcsname\relax
\typeout{** WARNING: IEEEtran.bst: No hyphenation pattern has been}%
\typeout{** loaded for the language `#1'. Using the pattern for}%
\typeout{** the default language instead.}%
\else
\language=\csname l@#1\endcsname
\fi
#2}}
\providecommand{\BIBdecl}{\relax}
\BIBdecl

\bibitem{sha48}
C.~E. Shannon, ``A mathematical theory of communication,'' \emph{Bell System
  Technical Journal}, vol.~27, pp. 379--423, Oct. 1948.

\bibitem{gal62}
R.~G. Gallager, ``Low-density parity-check codes,'' \emph{IRE Transactions on
  Information Theory}, vol.~8, no.~1, pp. 21--28, Jan. 1962.

\bibitem{lmss01}
M.~G. Luby, M.~Mitzenmacher, M.~A. Shokrollahi, and D.~A. Spielman, ``Improved
  low-density parity-check codes using irregular graphs,'' \emph{IEEE
  Transactions on Information Theory}, vol.~47, no.~2, pp. 585--598, Feb. 2001.

\bibitem{ru08}
T.~J. Richardson and R.~L. Urbanke, \emph{Modern coding theory}.\hskip 1em plus
  0.5em minus 0.4em\relax Cambridge University Press, 2008.

\bibitem{fz99}
A.~{Jim\'{e}nez Felstr\"{o}m} and {K. Sh. Zigangirov}, ``Time-varying periodic
  convolutional codes with low-density parity-check matrices,'' \emph{IEEE
  Transactions on Information Theory}, vol.~45, no.~6, pp. 2181--2191, Sept.
  1999.

\bibitem{psvc11}
A.~E. Pusane, R.~Smarandache, P.~O. Vontobel, and D.~J. {Costello, Jr.},
  ``Deriving good {LDPC} convolutional codes from {LDPC} block codes,''
  \emph{IEEE Transactions on Information Theory}, vol.~57, no.~2, pp. 835--857,
  Feb. 2011.

\bibitem{lscz10}
M.~Lentmaier, A.~Sridharan, D.~J. {Costello, Jr.}, and {K. Sh. Zigangirov},
  ``Iterative decoding threshold analysis for {LDPC} convolutional codes,''
  \emph{IEEE Transactions on Information Theory}, vol.~56, no.~10, pp.
  5274--5289, Oct. 2010.

\bibitem{kru11}
S.~Kudekar, T.~J. Richardson, and R.~L. Urbanke, ``Threshold saturation via
  spatial coupling: why convolutional {LDPC} ensembles perform so well over the
  {BEC},'' \emph{IEEE Transactions on Information Theory}, vol.~57, no.~2, pp.
  803--834, Feb. 2011.

\bibitem{ddja09}
D.~Divsalar, S.~Dolinar, C.~Jones, and K.~Andrews, ``Capacity-approaching
  protograph codes,'' \emph{IEEE Journal on Selected Areas in Communications},
  vol.~27, no.~6, pp. 876--888, Aug. 2009.

\bibitem{lmfc10b}
M.~Lentmaier, D.~G.~M. Mitchell, G.~P. Fettweis, and D.~J. {Costello, Jr.},
  ``Asymptotically good {LDPC} convolutional codes with {AWGN} channel
  thresholds close to the {Shannon} limit,'' in \emph{Proc. 6th International
  Symposium on Turbo Codes and Iterative Information Processing}, Brest,
  France, Sept. 2010.

\bibitem{tho03}
J.~Thorpe, ``Low-density parity-check ({LDPC}) codes constructed from
  protographs,'' Jet Propulsion Laboratory, Pasadena, CA, INP Progress Report
  42-154, Aug. 2003.

\bibitem{ips+12}
A.~R. Iyengar, M.~Papaleo, P.~H. Siegel, J.~K. Wolf, A.~{Vanelli-Coralli}, and
  G.~E. Corazza, ``Windowed decoding of protograph-based {LDPC} convolutional
  codes over erasure channels,'' \emph{IEEE Transactions on Information
  Theory}, vol.~58, no.~4, pp. 2303--2320, Apr. 2012.

\bibitem{lpf11}
M.~Lentmaier, M.~M. Prenda, and G.~Fettweis, ``Efficient message passing
  scheduling for terminated {LDPC} convolutional codes,'' in \emph{Proc. IEEE
  International Symposium on Information Theory}, St. Petersburg, Russia, Aug.
  2011.

\bibitem{kru12}
\BIBentryALTinterwordspacing
S.~Kudekar, T.~Richardson, and R.~Urbanke, ``Spatially coupled ensembles
  universally achieve capacity under belief propagation,'' 2012. [Online].
  Available: \url{http://arxiv.org/abs/1201.2999}
\BIBentrySTDinterwordspacing

\bibitem{kymp12}
S.~Kumar, A.~J. Young, N.~Macris, and H.~D. Pfister, ``A proof of threshold
  saturation for spatially-coupled {LDPC} codes on {BMS} channels,'' in
  \emph{Proc. Fiftieth Annual Allerton Conference}, Monticello, IL, Oct. 2012.

\end{thebibliography}
\end{document}